\documentclass{ckm}                 

\usepackage{txfonts}            
\usepackage{relsize}
\usepackage{graphicx}
\usepackage{epsfig}
\RequirePackage{xspace}

\confname{Workshop on the CKM Unitarity Triangle, IPPP Durham, April
  2003}

\title{Inclusive $|V_{ub}|$ measurements at \babar\ }

\author{A Sarti on behalf of the \babar\ Collaboration}

\address {Universit\`a di Ferrara, Dipartimento di Fisica and INFN, I-44100 Ferrara, Italy}

\def\Vub  {\ensuremath{|V_{ub}|}\xspace}
\def\BB      {\ensuremath{B\Bbar}\xspace} 
\def\babar{\mbox{\slshape B\kern-0.1em{\smaller A}\kern-0.1em
    B\kern-0.1em{\smaller A\kern-0.2em R}}}
\def\pep2{PEP-II}
\def\epem       {\ensuremath{e^+e^-}\xspace}
\def\invfb   {\ensuremath{\mbox{\,fb}^{-1}}\xspace}
\mathchardef\Upsilon="7107
\def\Y#1S{\ensuremath{\Upsilon{(#1S)}}\xspace}
\def\FourS {\Y4S}
\def\geant      {\mbox{\tt GEANT}\xspace}
\def\nub        {\ensuremath{\overline{\nu}}\xspace}
\def\jetset     {\mbox{\tt Jetset \hspace{-0.5em}7.4}\xspace}
\def\Bb      {\ensuremath{\Bbar}\xspace}
\def\Bu      {\ensuremath{B^+}\xspace}
\def\Bub     {\ensuremath{B^-}\xspace}
\def\piz   {\ensuremath{\pi^0}\xspace}
\def\KS    {\ensuremath{K^0_{\scriptscriptstyle S}}\xspace} 
 
\def\Dbar    {\kern 0.2em\overline{\kern -0.2em D}{}\xspace}

\def\mes        {\mbox{$m_{\rm ES}$}\xspace}

\def\BzBzb   {\ensuremath{\Bz {\kern -0.16em \Bzb}}\xspace}
\def\BpBm    {\ensuremath{\Bu {\kern -0.16em \Bub}}\xspace}
\def\Bz      {\ensuremath{B^0}\xspace}
\def\Bzb     {\ensuremath{\Bbar^0}\xspace}
\def\Dstarp  {\ensuremath{D^{*+}}\xspace}
\def\ssbar {\ensuremath{s\overline s}\xspace}
\def\BR         {{\ensuremath{\cal B}\xspace}}
\def\Bbar    {\kern 0.18em\overline{\kern -0.18em B}{}\xspace}

\def\ps   {\ensuremath{\rm \,ps}\xspace}

\newcommand {\breco}{\ensuremath{B_{reco}}}
\newcommand {\Bxclnu}{\ensuremath{\Bb \rightarrow X_c \ell \bar{\nu}}}
\newcommand {\Bxulnu}{\ensuremath{\Bb \rightarrow X_u \ell \bar{\nu}}}
\newcommand {\Bxuenu}{\ensuremath{\Bb \rightarrow X_u e \bar{\nu}}}

\newcommand {\lbar}{\ensuremath{\overline{\Lambda}}}
\newcommand {\lone}{\ensuremath{\lambda_1}}
\newcommand {\mX}{\ensuremath{m_{X}}}
\newcommand {\Bxlnu}{\ensuremath{\Bb \rightarrow X \ell \bar{\nu}}}
\newcommand {\rusl}{\ensuremath{R_{u/sl}}}
\newcommand{\gevcc}{\ensuremath{{\mathrm{\,Ge\kern -0.1em V\!/}c^2}}\xspace}
\newcommand{\gevc}{\ensuremath{{\mathrm{\,Ge\kern -0.1em V\!/}c}}\xspace}
\newcommand{\mevcc}{\ensuremath{{\mathrm{\,Me\kern -0.1em V\!/}c^2}}\xspace}
\newcommand{\gev}{\ensuremath{\mathrm{\,Ge\kern -0.1em V}}\xspace}
\newcommand{\mev}{\ensuremath{\mathrm{\,Me\kern -0.1em V}}\xspace}
\newcommand {\D}{\ensuremath{D}}
\def\Dstar   {\ensuremath{D^*}\xspace}

\newcommand {\BtoDstrlnu} {\ensuremath{\Bb \rightarrow \Dstar \ell \nu}}

\newcommand {\BtoDlnu} {\ensuremath{\Bb \rightarrow \D \ell \nu}}

\newcommand {\BtoDDstrpilnu} {\ensuremath{\Bb \rightarrow \D^{(*)} \pi \ell \nu}}

\newcommand{\mevc}{\ensuremath{{\mathrm{\,Me\kern -0.1em V\!/}c}}\xspace}

\def\allepsmx{0.770}
\renewcommand{\allepsmx}{73.3}
\def\allepsu{0.326}
\renewcommand{\allepsu}{34.2}

\begin{document}

\begin{abstract}
We present two inclusive measurements of  Cabibbo-Kobayashi-Maskawa (CKM) matrix element $|V_{ub}|$: one uses the lepton energy spectrum ($E_{l}$) and the other uses the invariant mass of the hadronic system (\mX) to discriminate signal (\Bxulnu) and background (\Bxclnu) events in \Bxlnu ~ transitions. Both analyses are based on data samples collected by the \babar\ detector at the PEP-II asymmetric-energy $B$ Factory at SLAC.
\end{abstract}

\maketitle


The element \Vub\ of the CKM matrix~\cite{ckm} plays a 
 central role in  tests  of the unitarity of this matrix:
it's extraction, based on tree level decays, gives results 
that are independent of new physics contributions.
We report the determination of \Vub\  from two different measurements of the
inclusive charmless semileptonic $B$ branching fraction, 
$\BR(\Bxulnu)$\footnote{Charge-conjugate  states are  implied 
 throughout this paper.},
 using $E_{l}$ spectrum (endpoint) and the 
\mX ~ spectrum on the recoil of fully reconstructed $B$ 
mesons respectively.\newline
The selection of \Bxulnu~ events is hampered by the 
presence of a large \Bxclnu\
background: $E_{l}$ and \mX\ spectra are used to discriminate 
the two different transitions. 
The endpoint analysis is sensitive to approximately 10\% of
 the $E_{l}$ spectrum while the acceptance for the 
\mX\ approach is larger: $\simeq70\%$ of the \mX\ spectrum
is selected by analysis cuts.
The extrapolation of the measured rates to the full phase
space introduces theoretical uncertainties\footnote{The results
 presented are depending on the
  parton-duality to which no error is assigned.}\cite{CKMyellow2002}.
Results also depend on the shape function (SF) modeling
of $b$ quark Fermi motion inside the $B$ meson.\newline
Both measurements are based on data recorded by the \babar\
detector~\cite{BBRD} at the \pep2\ asymmetric-energy \epem storage
ring at SLAC. The endpoint analysis data sample consists of 
about 23 million \BB\ pairs (21\invfb) collected at the \FourS\ 
resonance (ON-peak), with an additional sample of 2.6\invfb 
 recorded about 40\mev below the \FourS peak (OFF-peak), while 
the \mX~ measurement uses a data sample of about 88 million \BB\ 
pairs (82\invfb) ON-peak.\newline
 Monte Carlo (MC) simulations of the \babar\ 
detector based on \geant4~\cite{geant} are used to optimize 
selection criteria and to determine  signal  efficiencies   and
  background  shapes. To simulate  \Bxclnu~ transitions three
 models are employed: \BtoDstrlnu~ decay is modeled following a
 parametrization of form  factors based on HQET \cite{hqet};
 for \BtoDlnu~ decays and higher mass charm meson states 
$\Bb\to D^{(**)} \ell\nub$ the ISGW2 model \cite{ref:isgwtwo}~ is
 used; nonresonant decays, \BtoDDstrpilnu, are modeled
 according to a prescription by Goity and Roberts \cite{gr}.
 In the endpoint analysis the MC simulation of \Bxulnu~ events
 is based on the ISGW2 model: the hadrons $X_{u}$ are represented by 
single particles or resonances with masses up to 1.5\gevcc~ and
nonresonant contributions are not included. In the \mX\
analysis \Bxulnu~ transitions are simulated  with 
an hybrid model which is a mixture of resonant
and nonresonant components.  
The Fermi motion  of the  $b$  quark
inside the  $B$ meson is  implemented in the 
nonresonant component  using the SF parameterization
described in~\cite{ref:fazioneubert}, and the fragmentation is  handled by
\jetset~\cite{ref:jetset}.  

\section{Endpoint analysis}

\begin{figure}
\begin{center}
\hbox{\hskip-0.2cm \epsfig{file=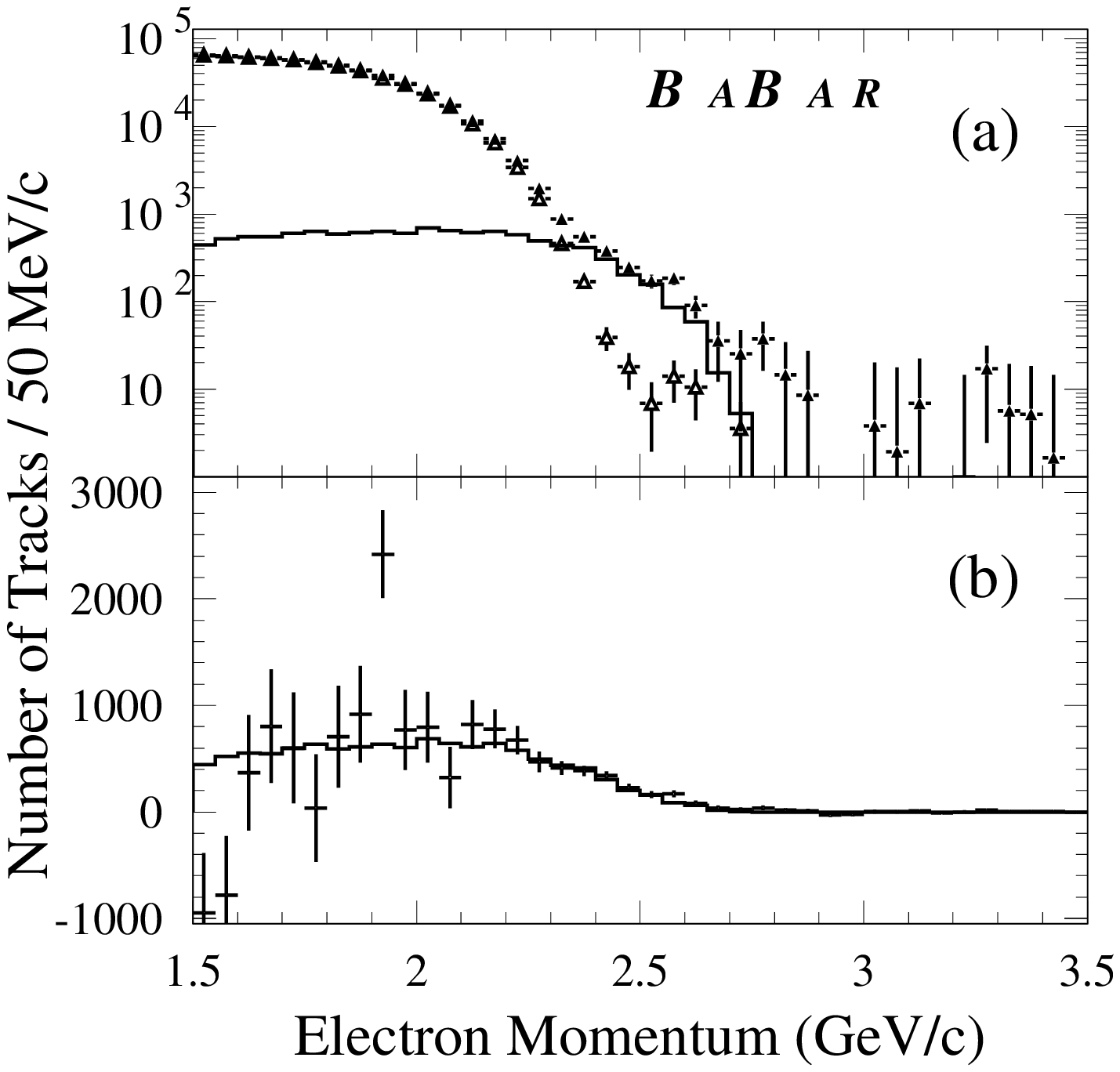,height=4.2cm} \hskip-0.2cm \epsfig{file=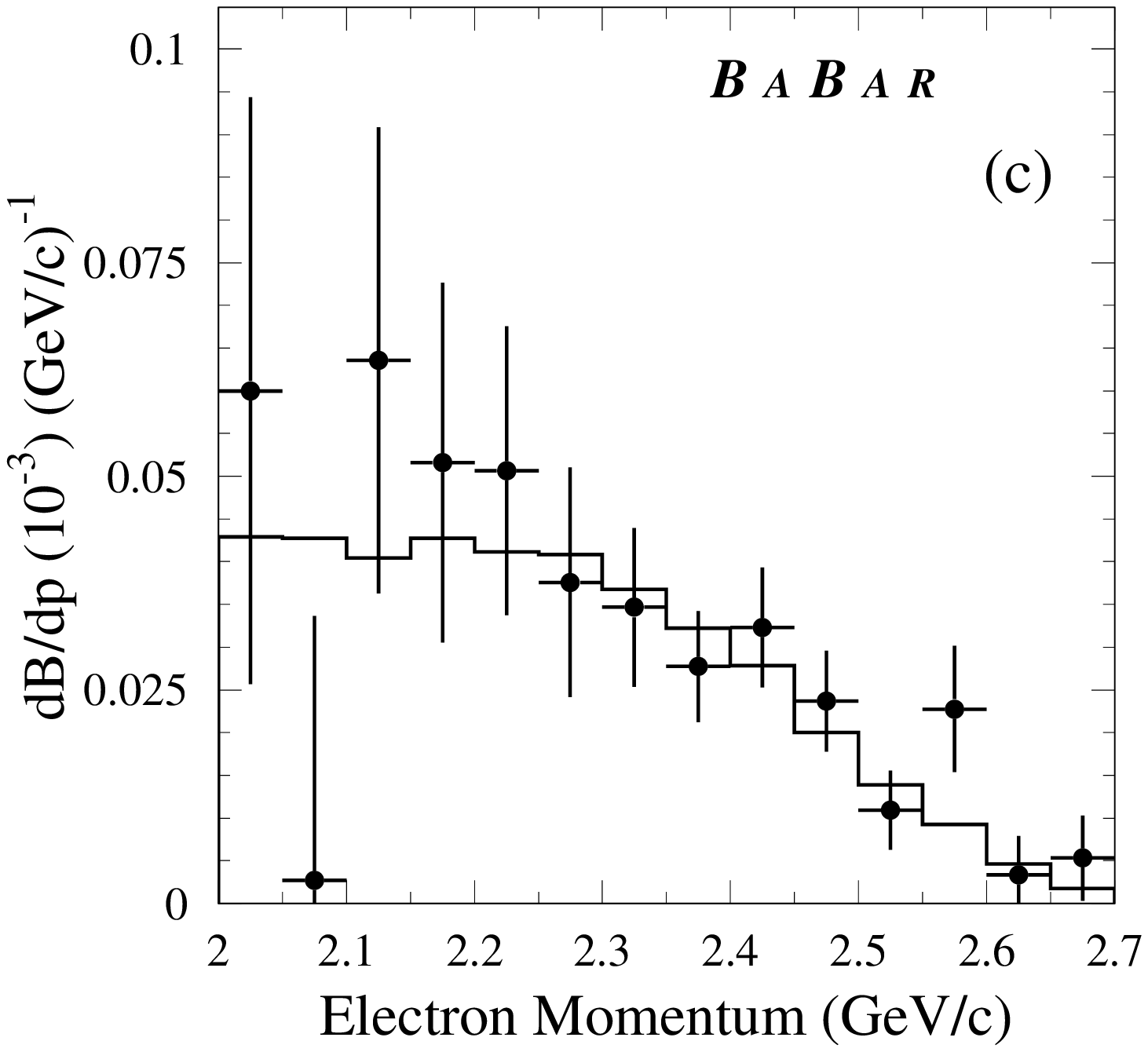,height=4.2cm}}
\caption{
Electron momentum spectrum in the $\FourS$ rest frame. 
(a) \textrm{ON}-peak data after continuum subtraction (solid triangles), and
MC predicted background from $B\bar{B}$ events ($B \nrightarrow X_u e \nu$) 
(open triangles). 
(b) \textrm{ON} peak data after subtraction of continuum and 
MC predicted $B \nrightarrow X_u e \nu$ backgrounds 
(data points with statistical errors). 
For comparison, the histograms show the expected signal spectrum from 
$B \to X_u e \nu$ decays.
(c) The differential rate $\textrm{\BR}(B \rightarrow X_u e \nu)$ 
as a function of 
the electron momentum. The data (statistical
errors only) are compared to the
prediction (solid line). 
}
\label{e1_spectrum}
\end{center}
\end{figure}

For this analysis, electron candidates are selected in the
momentum range from 1.5 to 3.5~\gevc in the \FourS\ rest frame with a solid 
angle defined by the electromagnetic calorimeter acceptance.\newline
The inclusive electron spectrum for charmless
 semileptonic $B$ decays, measured in the \FourS rest frame
in the momentum range of 2.3--2.6 \gevc, is used to extract $\BR(\Bxulnu)$.
 To suppress low-multiplicity QED processes and continuum processes
consisting of nonresonant $e^+e^-\to q\overline{q}$ 
production $(q = u, d, s, c)$ at least 
three charged tracks per event are required and a cut on the ratio of 
Fox-Wolfram moments $H_2/H_0<0.4$ \cite{foxw}~ is applied.
The missing momentum  four-vector $p_{miss} = p_i - p_{\breco} - p_X -
p_\ell$, where all momenta are measured in the laboratory frame and
$p_i$ refers to the four-momentum of the initial state of the
colliding beams,
 can be used to select semileptonic events:
 $|p_{miss}|$ is requested to be larger than 1~\gevc, 
 to point	
 into the detector fiducial volume and the angle
 between the electron candidate and the missing
 momentum is required to be greater than $\pi/2$.
 Candidate electrons are rejected if, when paired with an
opposite-sign electron, the invariant mass of the pair is consistent with
the $J/\psi$ mass ($3.05 < M_{e^+e^-} < 3.15$~\gevcc).
For the selection criteria described above, the detection efficiency  
for charmless semileptonic decays in the electron 
momentum interval of $1.5 - 2.6$~\gevc\ ranges from $\sim 0.4$ 
to $\sim 0.25$. \newline
The raw spectrum of the highest momentum electron after the subtraction 
of continuum background (determined from \textrm{OFF}
 resonance data sample) is 
shown in Fig.~\ref{e1_spectrum}a. 
Also shown are the MC predictions of the 
expected signal from \Bxuenu\ decays and background 
contributions from all other processes.  
The result of the subtraction of all backgrounds is shown 
in Fig.~\ref{e1_spectrum}b.
For a given interval in the electron momentum, the inclusive
partial branching ratio is calculated according to 
\begin{equation}
\hspace{1.0cm}\Delta \textrm{\BR} = \frac{N_{\mathrm{ON}}-N_{\mathrm{OFF}}-N_{B \nrightarrow X_u e \nu}}{2\epsilon N_{B\bar{B}}}(1+\delta_{\mathit{rad}}).
\end{equation}

\noindent 
Here $N_{\mathrm{ON}}$  refers to the number of 
electrons detected \textrm{ON}-peak and $N_{\mathrm{OFF}}$ refers to 
the fitted continuum background in a specified momentum interval,
$N_{{B} \nrightarrow X_u e \nu}$ is the 
background from $B\bar{B}$ decays, 
$\epsilon$ is the total efficiency for detecting a signal electron from 
$B \rightarrow X_u e\nu$ decays (including bremsstrahlung in detector 
material), and $\delta_{\mathit{rad}}$ accounts for the 
distortion of the electron spectrum due to final-state radiation. 
As the overall normalization the total number of produced $B\bar{B}$ events 
is used, $N_{B\bar{B}} = (22,630 \pm 19 \pm 362) \cdot 10^3$.\newline
The systematic error introduced by the efficiency estimation for the 
signal events is 5\%. The uncertainty in the continuum background subtraction
is 5\%. The error coming from the \BB~ background modeling
translates to a relative error of 3\%. 
Variations of the colliding beam energy 
introduce a systematic error in the $B\to X_c e\nu$
background subtraction (5\%). The total systematic error on 
the partial branching ratio measurement is $\sim 9\%$.\newline
The fully corrected
differential branching ratio as a function of the electron momentum 
is shown in Fig.~\ref{e1_spectrum}c. Integrating over the interval 
from 2.3 to 2.6~\gevc, we get:
\begin{equation}
\Delta \textrm{\BR} (B\rightarrow X_u e\nu)=(0.152\pm 0.014\pm 0.014)\cdot 10^{-3}
\end{equation}
\noindent 
To determine the charmless semileptonic branching fraction 
$\textrm{\BR} (B\rightarrow X_u e\nu)$ from the partial branching fraction 
$\Delta \textrm{\BR}(\Delta p)$, 
the fraction $f_u(\Delta p)$ of the spectrum that falls into 
the momentum interval $\Delta p$ is needed. The CLEO collaboration 
 has recently used the measurement 
of the inclusive photon spectrum from $b\to s\gamma$
 transitions \cite{CLEO-result} 
 to derive $f_u(\Delta p)$ for 
$B\rightarrow X_u e\nu$ transition.  They quote a value of 
$f_u(\Delta p) = 0.074 \pm 0.014 \pm 0.009$ for the interval $\Delta p$
from 2.3 to 2.6~\gevc.  Relying on the CLEO measurement, 
the result presented here 
translates into a total branching ratio of
$\textrm{\BR}(B\rightarrow X_u e\nu)=
(2.05 \pm 0.27_{exp} \pm 0.46_{f_u})\cdot 10^{-3}$.
We extract $|V_{ub}|$ from the measured inclusive charmless
semileptonic branching fraction with the relation in 
\cite{pdg2002} and the average $B$ lifetime 
of $\tau_B = 1.608 \pm 0.012\ps$~\cite{pdg2002}
and find 
\begin{equation}
|V_{ub}|= 
(4.43 \pm 0.29_{exp} \pm 0.25_{OPE} \pm 0.50_{f_u} \pm 0.35_{s\gamma}) \cdot 10^{-3}
\end{equation}
Here the first error is the combined statistical 
and systematic error, and the second refers to the
uncertainty on the extraction of $|V_{ub}|$ from relation in 
\cite{pdg2002}. 
The third one is taken from the CLEO analysis and 
is related to the experimental determination of $f_u$. 
The last error accounts for uncertainties related to assumption
that $b \to s \gamma$ spectrum can be used
 for shape function modeling
in $B\rightarrow X_u e\nu$ transition.

\section{\Vub\ measurement using the recoil of fully reconstructed B mesons}

\begin{figure}[t]
    \begin{centering} 
    \hbox{\hskip-0.cm \epsfig{file=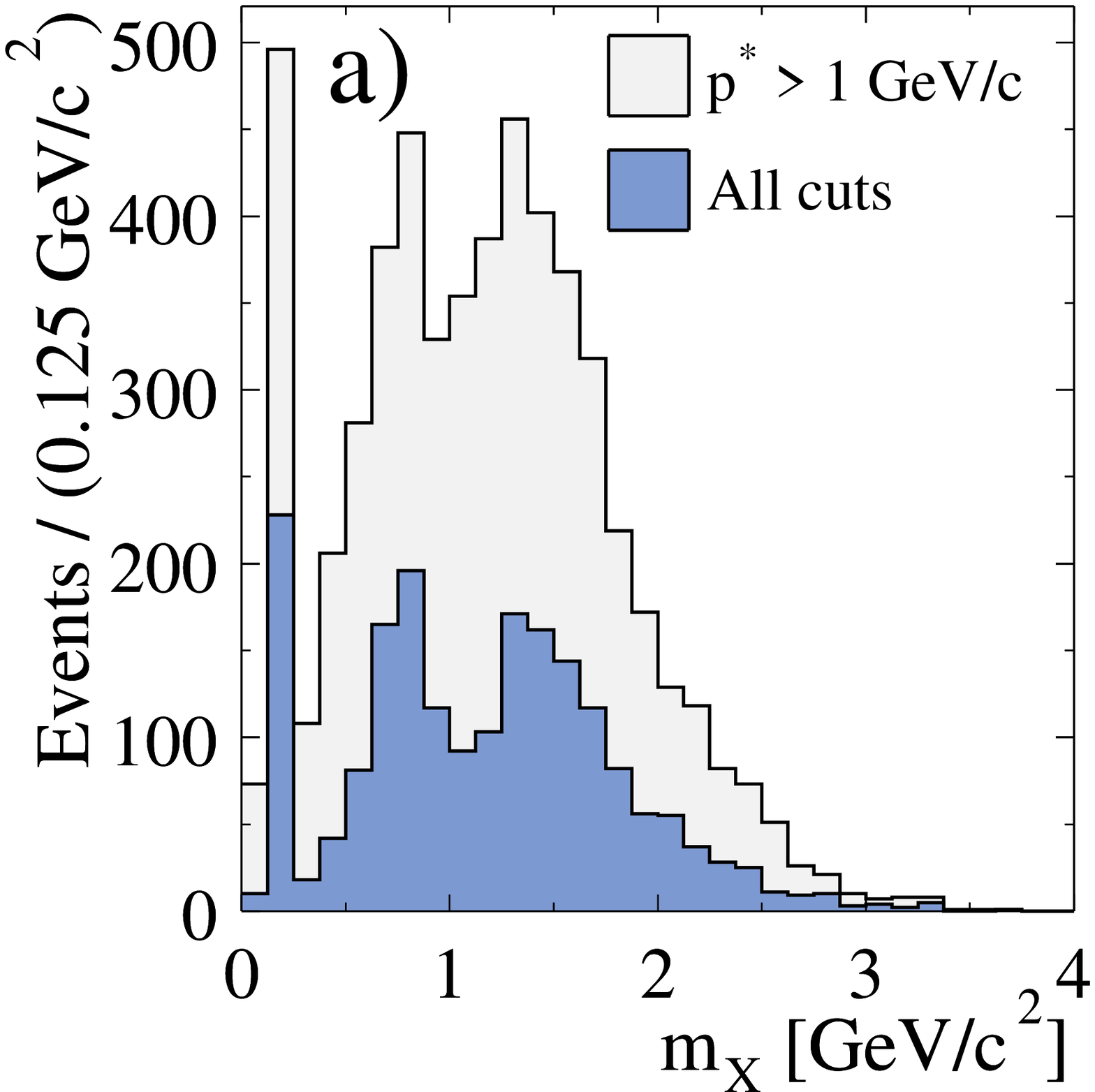,height=4.4cm} \hskip-0.3cm \epsfig{file=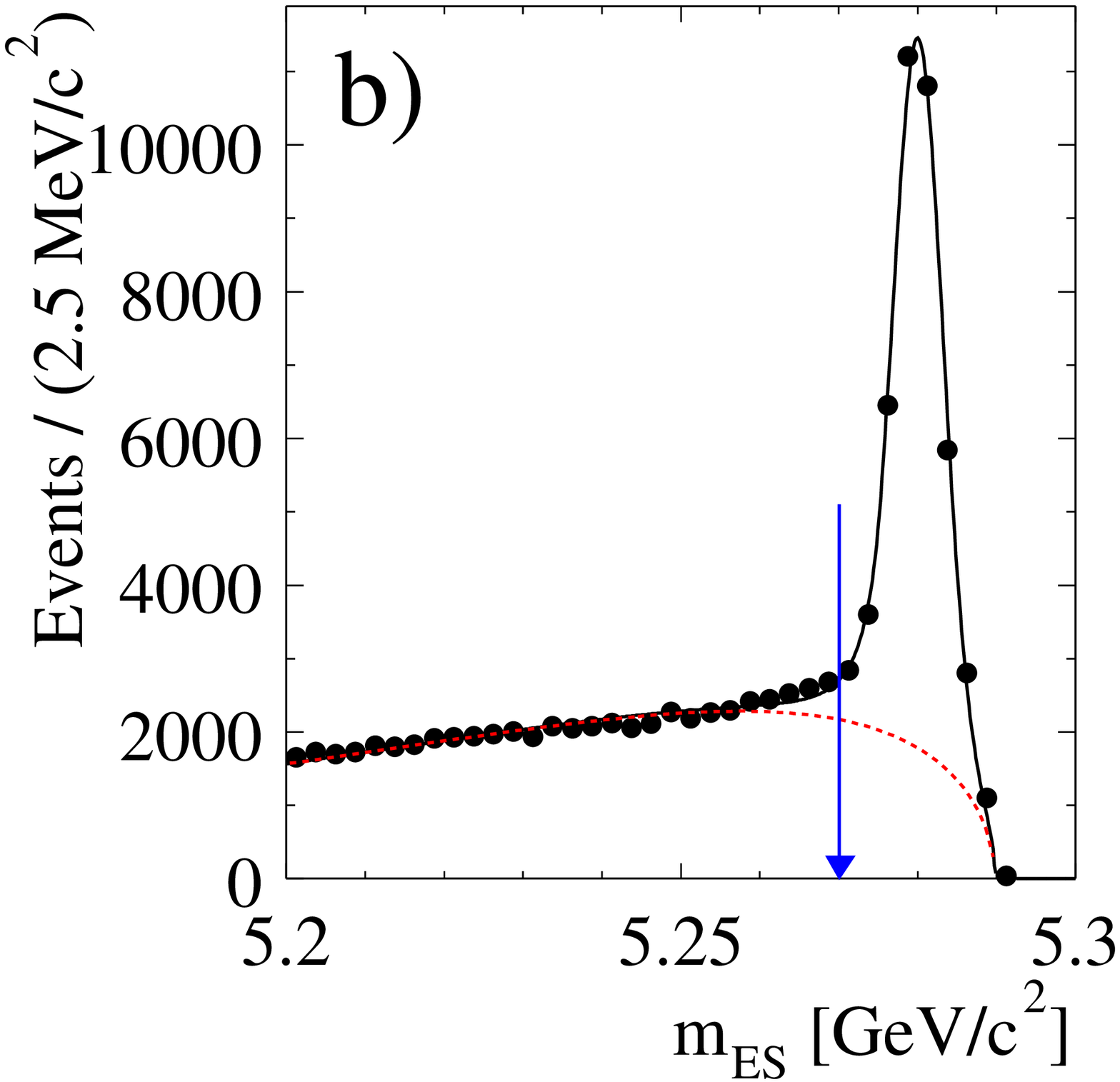,height=4.4cm}}
    \caption{
    a) Signal MC \mX\ distributions with the  requirement of a
      lepton with $p^* > 1\gevc$. Measured \mX\ distribution
      before and after all cuts.
    b)Fit to the $\mes$ distribution for the lepton sample
      with $p^* > 1\gevc$ in the recoil of a \breco\ candidate
    \label{fig:mxeff}}
   \end{centering}
   \end{figure}

\begin{figure}[t]
    \begin{centering}
     \hbox{\hskip -0.cm \epsfig{file=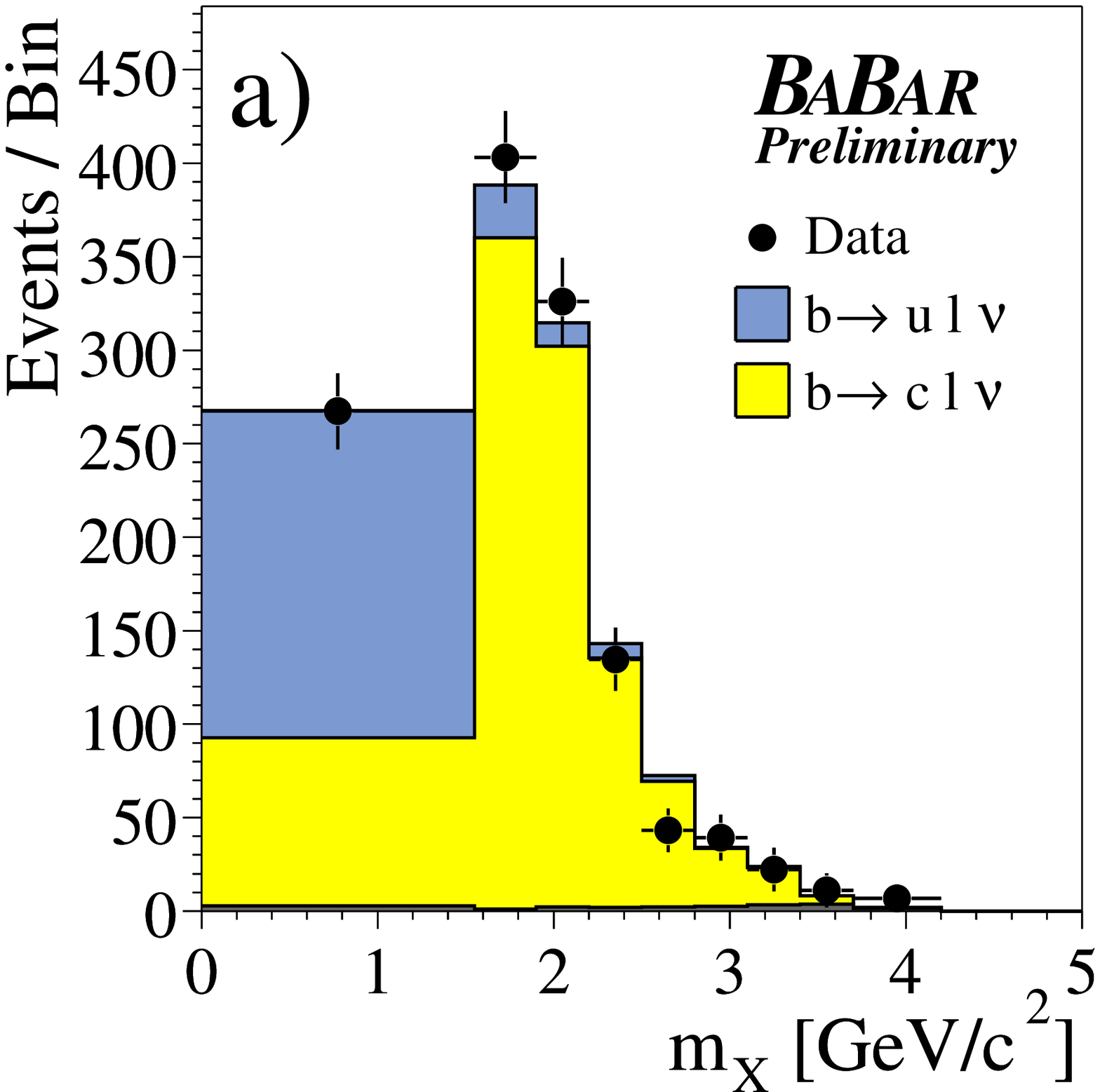,height=4.4cm} \hskip-0.3cm \epsfig{file=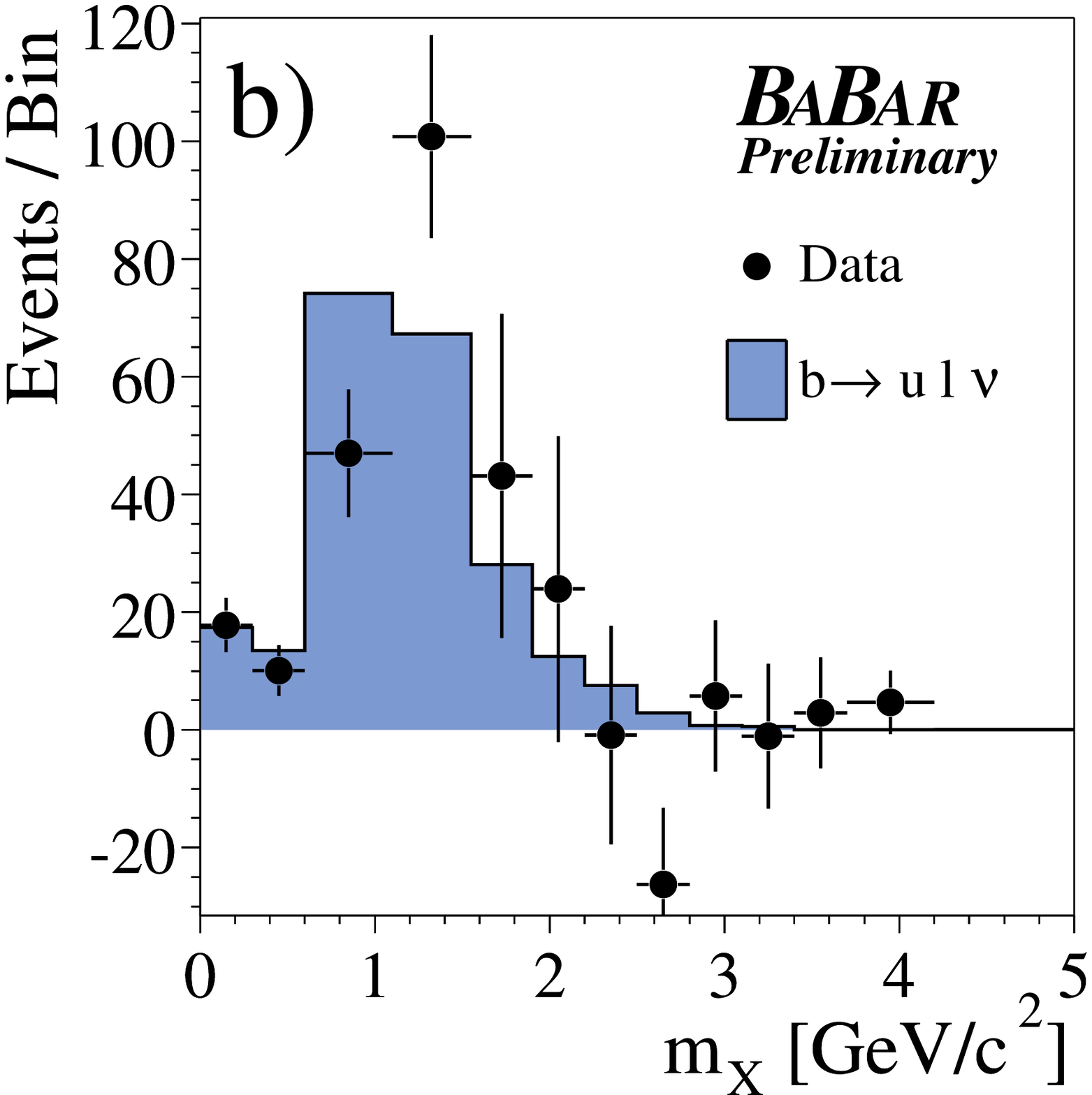,height=4.4cm} }
     \caption{The \mX\ distribution in \Bxlnu\ decays. a)
     Data (dots) and fit components. b) Background subtracted
     data and signal MC.  \label{fig:fitdata}}

   \end{centering}
   \end{figure}

This analysis is based on
\BB\ events in which one of the $B$ meson decays in a 
fully reconstructed hadronic final state ($B_{reco}$) and the 
other one is identified as 
decaying semileptonically by the presence of an electron or a muon.
The full reconstruction of one of the two $B$ mesons reduces the overall
efficiency, but allows to reconstruct both the neutrino and the
hadronic system ($X$), to determine the flavour
 and to separate charged and neutral $B$ mesons.
In order to reduce systematic uncertainties 
due to efficiency determination
 we extract the branching ratio $\rusl={\BR(\Bxulnu)/\BR(\Bxlnu)}$
after measuring the number of events with one identified lepton.\newline
To fully reconstruct a large sample of $B$ mesons, hadronic $B$ decays
of the type $B_{reco} \rightarrow  D^{(*)} Y$ are selected. $Y$ represents a
collection  of hadrons with  a total  charge of  $\pm 1$,  composed of
$n_1\pi^{\pm}\, n_2K^{\pm}\, n_3\KS\, n_4\piz$, where $n_1 + n_2 < 6$,
$n_3 < 3$,  and $n_4 < 3$.  The kinematic  consistency of a $B_{reco}$
candidate with a  $B$ meson decay is checked  using two variables, the
beam energy-substituted mass $\mes = \sqrt{s/4 -
\vec{p}^{\,2}_B}$ and the energy difference, 
$\Delta E = E_B - \sqrt{s}/2$. Here $\sqrt{s}$ refers to the total
energy in the \FourS center of mass frame, and $\vec{p}_B$ and $E_B$ denote
the momentum and energy of the $B_{reco}$ candidate in the same frame,
respectively.  In events  with more than one
reconstructed $B$ decay, the decay mode with highest purity is
selected. \newline
Semileptonic  \Bb\  decays, $\Bxlnu$,  recoiling  against the  \breco\
candidate  are identified  by an  electron  or muon  candidate with  a
minimum momentum ($p^*$) greater than $1\gevc$ in the \Bb rest frame.
 Correlation between the charge of the prompt
 leptons and the flavor of the $B_{reco}$ is imposed
($\Bz-\Bzb$ mixing rate is used to extract the
prompt lepton yield in case of neutral candidates). \newline
The hadron system $X$ in the decay \Bxlnu\ is made of charged
tracks  and neutral energy depositions
in the calorimeter that are not associated with
the \breco\ candidate and not identified as a lepton.
The mass of the hadronic system is determined by a
kinematic fit that imposes four-momentum conservation, the equality of
the masses of the two $B$ mesons, and forces $p_{miss}^2 = 0$.\newline
The selection of \Bxulnu\ decays is tightened by requiring exactly
one  charged lepton  with $p^*  >  1 \gevc$, charge  conservation
($Q_{X} + Q_\ell + Q_{\breco} =  0$), and  missing  mass consistent with zero 
($p_{miss}^{2} <  0.5 \gev^2/c^4$).  These criteria improve  the resolution in
\mX\ and  suppress the
dominant \Bxclnu\  decays, many of which  contain additional neutrinos
or   undetected  $K_L$. 
We suppress the $\Bzb\to\Dstarp\ell^-\overline{\nu}$ background   
with a  partial reconstruction  in  which only  the  slow  pion
from the  $\Dstarp\to \D\pi_s^+$ decay and the lepton are
reconstructed.
We veto events with charged or neutral kaons in the $X$
system to reduce the background from \Bxclnu\ decays.
The impact of the selection criteria on the \mX\ distribution is
illustrated on MC in Fig.~\ref{fig:mxeff}a.  
We determine \rusl\ from $N_u$, the observed number of $b\to u$ 
events, and $N_{sl}$, the number of events 
with at least one charged lepton: 
\begin{equation}
\rusl=
\frac{\BR(\Bxulnu)}{\BR(\Bxlnu)}=
\frac{N_u/(\varepsilon_{sel}^u \varepsilon_{\mX}^u)}{N_{sl}} 
\times \frac{\varepsilon_l^{sl} \varepsilon_t^{sl} } {\varepsilon_l^u \varepsilon_t^u }.
\label{eq:vubExtr}
\end{equation}
Here  $\varepsilon^u_{sel} = (\allepsu\pm0.6)\%$ is the
efficiency for selecting \Bxulnu\ decays with all analysis
requirements,  $\varepsilon^u_{\mX} = (\allepsmx\pm0.9)\%$ is the
fraction of signal events with $m_X < 1.55\gevcc$, 
$\varepsilon_l^{sl}/\varepsilon_l^u = 0.887\pm0.008$ corrects for the
difference in the efficiency due to the lepton momentum cut for
\Bxlnu\ and
\Bxulnu\ decays, and  $\varepsilon_t^{sl}/\varepsilon_t^u = 1.00 \pm 0.04$
accounts for a possible efficiency difference in the $B_{reco}$
reconstruction in events with \Bxlnu\ and \Bxulnu\ decays. \newline
We derive $N_{sl}$  from a fit to the \mes 
distribution shown in Fig.~\ref{fig:mxeff}b.
The fit uses an empirical description~\cite{argusf} of the combinatorial 
background from continuum and \BB\ events, together with a narrow
 signal~\cite{cry} peaked at the $B$ meson mass.
The  residual  background  in  $N_{sl}$  from  misidentified
leptons and semileptonic charm decays amounts to $6.8\%$ and
has been subtracted.
We obtain $N_u$ from   the \mX\ distribution with a $\chi^2$
fit to the sum of three contributions: signal, background
$N_{c}$ from
\Bxclnu, and a background of less than $1\%$ from other sources (misidentified
leptons, secondary $\tau$ and charm decays). \newline
 In each bin of the
\mX\ distribution, the combinatorial $B_{reco}$ background is
subtracted on the basis of a fit to the $\mes$ distribution. 
Figure~\ref{fig:fitdata}a shows the \mX\ distribution with 
the results of the fit superimposed.
 The fit reproduces well the data having a $\chi^2/dof=7.6/6$. 
In the fit, the first bin is chosen to contain all events with \mX\ less than
$1.55\gevcc$  while the other bins are chosen in order to separate
the contribution from  each resonant $\Bxclnu$
mode. The \mX\ cut, set at $1.55\gevcc$, has been optimized 
minimizing the total error.
Figure~\ref{fig:fitdata}b shows the \mX\ distribution
after background subtraction with finer binning.
 Table ~\ref{breakdown} summarizes the results of fits
 with different requirements on \mX,
 for electrons and muons, for neutral and charged \breco\ candidates, 
and for different ranges of the $B_{reco}$ purity, $\cal P$ .
The results are all consistent within the 
uncorrelated statistical errors.\newline
We have performed extensive studies to determine systematic
uncertainties.
 ~We use events with charged and neutral kaons in the
recoil of the \breco\ candidate as a control sample to assess that  the
background from \Bxclnu\  events is properly described.
The relative systematic error ($\Delta_r$) due to the selection criteria
related to the reconstruction of particles in the event
is $\Delta_r=8.5\%$.
The uncertainty of
the $B_{reco}$ combinatorial background subtraction is estimated by
varying the signal shape function ($\Delta_r=3.8\%$).  The
impact of the binning  is studied by changing the binning for $\mX >
1.55\gevcc$ ($\Delta_r=2.9\%$). The branching fractions
 of $B\to D^{(*,**)}\ell\nu$ and
of inclusive and exclusive $D$ mesons decays are varied within
the world average uncertainties~\cite{pdg2002} ($\Delta_r=4.4\%$). The
limited amount of simulated events causes an uncertainty $\Delta_r=5.1\%$.
The uncertainty in the hadronization of the final state of \Bxulnu\
events is determined by measuring $\rusl$ in bins of charged and
neutral multiplicities and performing the fit using only the nonresonant
signal model instead of the hybrid model($\Delta_r=3.0\%$). We also vary
 the branching
fractions for charmless semileptonic $B$ decays by one standard
deviation~\cite{pdg2002}($\Delta_r=2.8\%$).  The fraction of signal events with \ssbar\
contents is varied by $100\%$ for the exclusive component and by
$30\%$ for the inclusive one \cite{Althoff:1984iz}($\Delta_r=3.7\%$).
In the determination of $\varepsilon_{sel}^{u}$ and
$\varepsilon_{\mX}^{u}$ we allow the nonperturbative parameters to
vary according to $\lbar = 0.480 \pm 0.120 \gev$ and $\lone =
-0.300\pm0.105\gev^2$, obtained by scaling the results
in~\cite{ref:cleomxhad} to  ${\cal O}(1/m_b^2, \alpha_s)$ 
in order to match the nonresonant MC generator~\cite{ref:fazioneubert}.
  We take into account the correlation of $-0.8$ between 
\lbar\ and \lone($\Delta_r=17.5\%$).\newline
We combine the errors related to the 
detector and the signal and background
modeling errors quadratically into the systematic error and obtain
$\rusl = 0.0197 \pm 0.0027 \pm 0.0023 \pm 0.0034$,
where the errors are statistical, systematic, and theoretical related
to the efficiency determination and extrapolation to the full \mX\
range respectively.  Combining the ratio
\rusl\ with the measured inclusive semileptonic branching fraction of
$\BR(\Bxlnu) = (10.87 \pm 0.18(stat) \pm
0.30(sys))\%$~\cite{Aubert:2002uf}, we obtain
$\BR(\Bxulnu) = (2.14 \pm 0.29 \pm 0.26 
\pm 0.37)\times10^{-3}$.
Using the relation in ~\cite{pdg2002} and the average $B$ lifetime of $\tau_B = 1.608 \pm 0.016\ps$~\cite{pdg2002} we find 
\begin{equation}
\Vub = (4.52
\pm0.31
\pm0.27
\pm0.40
\pm0.25)\times 10^{-3}
\end{equation}
The first error is statistical, the second refers to the experimental
systematic uncertainty, the third gives the theoretical uncertainty
on the extrapolation of $R_{u/sl}$ to the full \mX\ range, 
and the last error combines quadratically the perturbative and
nonperturbative uncertainties in the extraction of \Vub\ from the
total decay rate.\begin{table}
\begin{center}
\caption{Fit results for several data samples.}
\scriptsize
\vspace{0.03in}
\begin{tabular}{|lrrrr|} \hline 
Sample  &$N_{sl}$\,\,&$N_u$\quad\,& $N_{c}$ &$R_{u/sl}$\quad(\%)\\ 
\hline
$\mX<1.55\gevcc$     &$32210\pm233$   &$167\pm21$   &$99\pm6$   &$1.97\pm0.25$\\
$\mX<1.40\gevcc$     &$32210\pm233$   &$134\pm19$   &$64\pm4$   &$1.77\pm0.25$\\
$\mX<1.70\gevcc$     &$32210\pm233$   &$191\pm26$   &$170\pm11$ &$2.11\pm0.29$\\
\hline                                                              
neutral $B_{reco}$   &$11582\pm133$   &$76\pm13$    &$21\pm3$  &$2.46\pm0.43$\\
charged $B_{reco}$   &$20583\pm191$   &$91\pm16$    &$77\pm5$  &$1.68\pm0.30$\\
\hline                                                              
Electrons            &$18261\pm173$   &$99\pm15$    &$48\pm4$  &$2.26\pm0.35$\\
Muons                &$13934\pm157$   &$67\pm14$    &$47\pm4$  &$1.66\pm0.36$\\
\hline                                                              
${\cal P}>80\%$      &$4491\pm68$     &$19\pm7$     &$13\pm2$  &$1.59\pm0.56$\\
$50\%<{\cal P}<80\%$ &$13298\pm141$   &$65\pm13$    &$46\pm3$  &$1.85\pm0.37$\\
${\cal P}<50\%$      &$14122\pm170$   &$82\pm15$    &$38\pm3$  &$2.21\pm0.40$\\
\hline
\end{tabular}
\label{breakdown}
\end{center}
\end{table}

\section{Conclusions}

Two different approaches for the extraction of \Vub\ CKM matrix
 element  have been presented. The analysis based on \mX\ gives 
currently the most precise determination of \Vub.

This is primarily due to specific advantages
of this technique: large phase-space acceptance and high 
purity of the sample (signal over background ratio $\sim1.7$).
 The two results are consistent 
 and they are in agreement with previous inclusive 
measurements~\cite{Barate:1998vv}.


\begin{thebibliography}{9}

\bibitem{ckm}
N.~Cabibbo, 
Phys.\ Rev.\ Lett.\  {\bf 10}, 531 (1963); M. Kobaya\-shi and T. Maskawa, 
Prog.\ Th.\ Phys. {\bf 49}, 652 (1973).


\bibitem{CKMyellow2002}
M.~Battaglia {\it et al.},
``The CKM matrix and the unitarity triangle,''
arXiv:hep-ph/0304132, 100 (2002)


\bibitem{BBRD}    
The \babar\ Collaboration, B.\ Aubert {\em et al}., Nucl. Instrum. Methods. 
{\bf A479}, 1 (2002).


\bibitem{geant}
S.~Agostinelli {\it et al.}  [GEANT4 Collaboration],
SLAC-PUB-9350, CERN-IT-2002-003 (2002)

\bibitem{hqet}
I.I.\ Bigi, M.\ Shifman, and N.G.\ Uraltsev, Annu. Rev. Nucl. Part. Sci. 
{\bf 47}, 591 (1997); 
J.~E.~Duboscq {\it et al.}  [CLEO Collaboration],
Phys.\ Rev.\ Lett.\  {\bf 76}, 3898 (1996).

\bibitem{ref:isgwtwo}
D.~Scora and N.~Isgur,
Phys.\ Rev.\ D {\bf 52}, 2783 (1995).
%
\bibitem{gr}
J.L.\ Goity and W.\ Roberts, Phys. Rev. {\bf D51}, 3459 (1995).
%

\bibitem{ref:fazioneubert} 
F.~De Fazio and M.~Neubert,
JHEP {\bf 9906}, 017 (1999).
\bibitem{ref:jetset} 
T.~Sj\"ostrand, Comput. Phys. Commun.~{\bf 82}, 74 (1994).

\bibitem{foxw}
G.C.\ Fox and S.\ Wolfram, Phys. Rev. Lett. {\bf 41}, 1581 (1978). 
%
\bibitem{CLEO-result}
The CLEO Collaboration, A.\ Bornheim {\em et al}., hep-ex/0202019 (2002).
%
\bibitem{Aubert:2002uf}
B.~Aubert {\it et al.}  [BABAR Collaboration],
Phys.\ Rev.\ D {\bf 67}, 031101 (2003).




\bibitem{argusf}
H.~Albrecht {\it et al.}  [ARGUS Collaboration],
Z.\ Phys.\ C {\bf 48}, 543 (1990).

\bibitem {cry}
T.~Skwarnicki  [Crystal Ball Collaboration],
DESY F31-86-02.


\bibitem{pdg2002}
K.~Hagiwara {\it et al.}  [Particle Data Group Collaboration],
Phys.\ Rev.\ D {\bf 66}, 010001 (2002).


\bibitem{Althoff:1984iz}
M.~Althoff {\it et al.}  [TASSO Collaboration],
Z.\ Phys.\ C {\bf 27}, 27 (1985).
W.~Bartel {\it et al.}  [JADE Collaboration],
Z.\ Phys.\ C {\bf 20}, 187 (1983).
V.~Luth {\it et al.},
Phys.\ Lett.\ B {\bf 70}, 120 (1977).

\bibitem{ref:cleomxhad}
D.~Cronin-Hennessy {\it et al.} [CLEO Collaboration], 
Phys.\ Rev.\ Lett.\ 87:251808,\ 2001.



\bibitem{Barate:1998vv}
R.~Barate {\it et al.}  [ALEPH Collaboration],
Eur.\ Phys.\ J.\ C {\bf 6}, 555 (1999);
%
M.~Acciarri {\it et al.}  [L3 Collaboration],
Phys.\ Lett.\ B {\bf 436}, 174 (1998);
%
P.~Abreu {\it et al.}  [DELPHI Collaboration],
Phys.\ Lett.\ B {\bf 478}, 14 (2000);
%
B.~H.~Behrens {\it et al.}  [CLEO Collaboration],
Phys.\ Rev.\ D {\bf 61}, 052001 (2000);
%
G.~Abbiendi {\it et al.}  [OPAL Collaboration],
Eur.\ Phys.\ J.\ C {\bf 21}, 399 (2001);
%
A.~Bornheim {\it et al.}  [CLEO Collaboration],
Phys.\ Rev.\ Lett.\  {\bf 88}, 231803 (2002).
 
\end{thebibliography}
\end{document}